\documentclass{jps-cp}
\usepackage{txfonts} %Please comment out this line unless the txfonts package is availabe in your LaTeX system.

\title{Time-like Gravitational 
Formfactors and Shear 
Viscosity 
}

\author{Oleg \textsc{Teryaev}$^{1,2}$} 
\inst{$^{1}$Joint Institute for Nuclear Research,  Dubna 141980, Russia \\
$^{2}$Dubna State University, Dubna 141980, Russia}

\email{teryaev@jinr.ru}

\recdate{February 15, 2022}

\abst{The gravitational formfactor similar to shear viscosity 
is identified. In the time-like region it corresponds to the contribution of exotic hybrid meson. The exotic quantum numbers may be considered as a counterpart of dissipation in crossed channel. The ratio of viscosity to entropy density is estimated. The smallness due to holographic bound corresponds to the relative smallness of coupling of hybrid meson and the intrinsic transverse momentum, being the dynamical counterpart of temperature.   
}

\kword{gravitational formfactors, viscosity, hybrid, bound, duality}

\begin{document}
\maketitle

\section{Introduction}
The gravitational formfacors \cite{Pagels:1966zza}  accumulate the important infromation on 
hadronic structure \cite{Ji:1996ek,Polyakov:2018zvc}  
and  behaviour of hadrons in gravitational field \cite{Teryaev:1999su,Teryaev:2016edw}.

The gravitational form factors \cite{Teryaev:2016edw}, related to pressure, may be also considered in the time-like region, and this opportunity was
recently explored in the exclusive production of pion pairs in the collisions of real and virtual photons to get the relevant information for pion  \cite{Kumano:2017lhr}.

There is a possibility \cite{Teryaev:2019ale} to attribute to gravitational formfactors such seemingly different quantity as shear viscosity. Here we discuss this opportunity in more detail and perform the estimate of viscosity showing the possible  counterparts to its smallness established in holographic approach \cite{Policastro:2001yc,Kovtun:2004de,Grozdanov:2016fkt}.

\section{Viscosity and formfactors} 

Let us estimate the haddronic analog of viscosity term in energy momentum tensor
\begin{eqnarray}
\eta  \frac {\partial v^\nu}{\partial x_\mu} \to \frac {P^\nu \Delta^\mu}{M} \sim E (t) P_\nu \Delta_\mu, \label{eta}
\end{eqnarray}
where $E(t)$ is the formfactor to be discussed below. The proportionality 
constant is not affecting the ratio to be estimated and requires the special investigation while the states defining the formfactor will later appear to be different from each other.

Here we assume for this estimate that hadron "velocity" $v_\mu=P_\mu/M$. Let us stress that hadron should be considered here  as a sort of "liquid" rather than spherically symmetric \cite{Polyakov:2018zvc} object. Such velocity has a correct normalization but  does not have a dependence on the transverse coordinate, which may be restored, say, by modification   
\begin{eqnarray}
\label{v}
 v^\mu = \frac {P^\mu + a (t) \Delta^\mu}{\sqrt{M^2 - a^2(t) t }}, 
\end{eqnarray}
which is neglected in estimate (\ref{eta}). The transverse derivative is  
estimated as 
\begin{eqnarray}
\label{un}
 \frac{\partial }{\partial x_\mu} \to i \Delta^\mu. 
\end{eqnarray}
The similar estimate for the standard term in liquid energy-momentum tensor looks like 
\begin{eqnarray}
\label{s}
 (e+p) v^\nu v^\mu \to T s \frac {P^\nu P^\mu}{M^2} \sim A (t) P^\nu P^\mu.
\end{eqnarray}
Comparing (\ref{eta}, \ref{s}) one get
\begin{eqnarray}
\label{R}
\frac{\eta}{s} \sim \frac {E(t)}{A(t)}\cdot \frac{T}{M}.
\end{eqnarray}
The smallness of this ratio may be due to the smallness of $E$ formfactor (which cannot appear in the conserved energy momentum tensor due to equivalence principle, but may appear for quarks separately, as it will be discussed in the next section) and hadronic "temperature" which may be substituted by average transverse momentum  \cite{Cleymans:2011im,Cleymans:2013rfq}.

These properties can be better understood when considering  time-like 
 region.

\section{Time-like region}

The gravitational form factors \cite{Teryaev:2016edw}, related to pressure, may be also considered in the time-like region, and this opportunity was
recently explored in the exclusive production of pion pairs in the collisions of real and virtual photons to get the relevant information for pion  \cite{Kumano:2017lhr}.

It will be very interesting to extend this analysis for the production of $\pi \eta$ pairs with exotic quantum numbers $J^{PC}=1^{-+}$, being the natural generalization of the production of exotic hybrid mesons which were considered earlier \cite{Anikin:2006du}.
The corresponding Generalized Distribution Amplitude (GDA) is the straightforward counterpart of (\ref{eta}), the crossing changing 
$\Delta$ to the total momentum of the pair and $P$ to the relative momentum of the mesons. 
\begin{eqnarray}
\langle \pi \eta (P, \Delta) |T_i^{\alpha \nu}|0 \rangle_{\mu^2}= E_i (s, \mu^2) P^{\alpha} \Delta^{\nu}
\end{eqnarray} 
For the vector  meson $P^\alpha/M$ can be also understood as its polarization vector. This clearly shows that mesons should be different, as the parity would be negative otherwise. 

For the GPD channel this shows that the $\pi \to \eta$ {\it transition} distribution amplitude should be considered, being the counterpart of {\it dissipation} and T-oddness. Note also that the interplay between T symmetry in GPD channel and C symmetry in GDA channel is essential to the application of Radon transform and may be considered as a manifestation of CPT symmetry in P-conserving case \cite{Teryaev:2001qm}.

Needless to say, that the total average viscosity of quarks and gluons should be zero, which is a natural generalization of nullification of exotic hybrid meson coupling \cite{Anikin:2006du,Anikin:2004vc} and, as it can be added now, is in full agreement with equivalence principle.

The smallness of holographic bound can be also attributed to the smallness of hybrid coupling  \cite{Anikin:2006du,Anikin:2004vc} 
($\sim 50 MeV$). One should expect that in (\ref{R}) its (small) ratio 
to that of typical C-even meson (like $\sigma$) should enter.

\section{Dimensionful expressions} 

It is instructive to consider the expression for the viscosity in the case when units different from $\hbar=k_B=1$ are used. In this case 
(\ref{un}) takes the form 
\begin{eqnarray}
\label{unh}
\hbar  \frac{\partial }{\partial x_\mu} \to i \Delta^\mu. 
\end{eqnarray}
 Consequently, (\ref{R}) looks like 
\begin{eqnarray}
\label{Rh}
\frac{\eta}{s} \sim \hbar \frac {E(t)}{A(t)}\cdot \frac{T}{M}.
\end{eqnarray}
If temperature is measured in the units different from $MeV$
the ratio $T/M$ is not dimensionless anymore. We can restore the dimensionless factor by multiplying the numerator and denominator by $k_B$: 
\begin{eqnarray}
\label{Rhk}
\frac{\eta}{s} \sim \frac {\hbar}{k_B} \cdot \frac {E(t)}{A(t)}\cdot \frac{k_B T}{M},
\end{eqnarray}
 resulting in the expression similar to \cite{Policastro:2001yc,Kovtun:2004de,Trachenko:2020ktm}.

Finally, when choice $c=1$ is also dropped, the estimate looks like
\begin{eqnarray}
\label{Rhkc}
\frac{\eta}{s} \sim \frac {\hbar}{k_B} \cdot \frac {E(t)}{A(t)}\cdot \frac{k_B T}{M c^2}.
\end{eqnarray}

\section{Conclusions}

We observe (somehow unexpectedly) that such a notion as shear viscosity may appear also in hadronic structure. The relevant formfactor 
is absent for the conserved energy-momentum tensor but may appear for quarks separately. The time-like region is of special interest as the 
properties of dissipation and T-oddness correspond there to the production of exotic hybrid mesons. Investigation of these effects using BELLE data similarly to \cite{Kumano:2017lhr} are rather promising.

Besides the relation of rather diverse phenomena of dissipation and exotics production, one should stress here that the new link between physics of hadronic and heavy-ion collisions appears. This can be of special interest in the studies \cite{Arbuzov:2020cqg} involving the NICA complex at JINR. 

\section{Acknowledgments}

I am indebted to S. Kumano and V.I. Zakharov for useful discussions and comments.

The work was supported by Russian Science Foundation Grant 21-12-00237.


\begin{thebibliography}{9}




%\cite{Pagels:1966zza}
\bibitem{Pagels:1966zza}
H.~Pagels,
``Energy-Momentum Structure Form Factors of Particles,''
Phys. Rev. \textbf{144}, 1250-1260 (1966)
doi:10.1103/PhysRev.144.1250
%124 citations counted in INSPIRE as of 14 Feb 2022
%\cite{Ji:1996ek}
\bibitem{Ji:1996ek}
X.~D.~Ji,
``Gauge-Invariant Decomposition of Nucleon Spin,''
Phys. Rev. Lett. \textbf{78}, 610-613 (1997)
doi:10.1103/PhysRevLett.78.610
[arXiv:hep-ph/9603249 [hep-ph]].
%1951 citations counted in INSPIRE as of 14 Feb 2022
%\cite{Polyakov:2018zvc}
\bibitem{Polyakov:2018zvc}
M.~V.~Polyakov and P.~Schweitzer,
``Forces inside hadrons: pressure, surface tension, mechanical radius, and all that,''
Int. J. Mod. Phys. A \textbf{33}, no.26, 1830025 (2018)
doi:10.1142/S0217751X18300259
[arXiv:1805.06596 [hep-ph]].
%144 citations counted in INSPIRE as of 14 Feb 2022

%\cite{Teryaev:1999su}
\bibitem{Teryaev:1999su}
  O.~V.~Teryaev,
  ``Spin structure of nucleon and equivalence principle,''
  hep-ph/9904376.
  %%CITATION = HEP-PH/9904376;%%
  %95 citations counted in INSPIRE as of 18 Aug 2019

%\cite{Teryaev:2016edw}
\bibitem{Teryaev:2016edw}
O.~V.~Teryaev,
``Gravitational form factors and nucleon spin structure,''
Front. Phys. (Beijing) \textbf{11}, no.5, 111207 (2016)
doi:10.1007/s11467-016-0573-6
%33 citations counted in INSPIRE as of 14 Feb 2022


%\cite{Kumano:2017lhr}
\bibitem{Kumano:2017lhr}
  S.~Kumano, Q.~T.~Song and O.~V.~Teryaev,
  ``Hadron tomography by generalized distribution amplitudes in pion-pair production process $\gamma^* \gamma \rightarrow \pi^0 \pi^0 $ and gravitational form factors for pion,''
  Phys.\ Rev.\ D {\bf 97} (2018)  014020
  doi:10.1103/PhysRevD.97.014020
  [arXiv:1711.08088 [hep-ph]].
  %%CITATION = doi:10.1103/PhysRevD.97.014020;%%
  %17 citations counted in INSPIRE as of 31 Jan 2019


%\cite{Teryaev:2019ale}
\bibitem{Teryaev:2019ale}
O.~Teryaev,
``Shear forces and tensor polarization,''
PoS \textbf{DIS2019}, 240 (2019)
doi:10.22323/1.352.0240
%1 citations counted in INSPIRE as of 14 Feb 2022

%\cite{Policastro:2001yc}
\bibitem{Policastro:2001yc}
G.~Policastro, D.~T.~Son and A.~O.~Starinets,
``The Shear viscosity of strongly coupled N=4 supersymmetric Yang-Mills plasma,''
Phys. Rev. Lett. \textbf{87}, 081601 (2001)
doi:10.1103/PhysRevLett.87.081601
[arXiv:hep-th/0104066 [hep-th]].
%1515 citations counted in INSPIRE as of 14 Feb 2022

%\cite{Kovtun:2004de}
\bibitem{Kovtun:2004de}
P.~Kovtun, D.~T.~Son and A.~O.~Starinets,
``Viscosity in strongly interacting quantum field theories from black hole physics,''
Phys. Rev. Lett. \textbf{94}, 111601 (2005)
doi:10.1103/PhysRevLett.94.111601
[arXiv:hep-th/0405231 [hep-th]].
%2585 citations counted in INSPIRE as of 14 Feb 2022

%\cite{Grozdanov:2016fkt}
\bibitem{Grozdanov:2016fkt}
S.~Grozdanov and A.~O.~Starinets,
``Second-order transport, quasinormal modes and zero-viscosity limit in the Gauss-Bonnet holographic fluid,''
JHEP \textbf{03}, 166 (2017)
doi:10.1007/JHEP03(2017)166
[arXiv:1611.07053 [hep-th]].
%79 citations counted in INSPIRE as of 14 Feb 2022

%\cite{Cleymans:2011im}
\bibitem{Cleymans:2011im}
J.~Cleymans, G.~I.~Lykasov, A.~S.~Sorin and O.~V.~Teryaev,
``Dynamical and thermal descriptions in parton distribution functions,''
Phys. Atom. Nucl. \textbf{75}, 725-728 (2012)
doi:10.1134/S1063778812060099
[arXiv:1104.0620 [hep-ph]].
%10 citations counted in INSPIRE as of 14 Feb 2022



%\cite{Cleymans:2013rfq}
\bibitem{Cleymans:2013rfq}
J.~Cleymans, G.~I.~Lykasov, A.~S.~Parvan, A.~S.~Sorin, O.~V.~Teryaev and D.~Worku,
``Systematic properties of the Tsallis Distribution: Energy Dependence of Parameters in High-Energy p-p Collisions,''
Phys. Lett. B \textbf{723}, 351-354 (2013)
doi:10.1016/j.physletb.2013.05.029
[arXiv:1302.1970 [hep-ph]].
%113 citations counted in INSPIRE as of 14 Feb 2022




%\cite{Anikin:2006du}
\bibitem{Anikin:2006du}
  I.~V.~Anikin, B.~Pire, L.~Szymanowski, O.~V.~Teryaev and S.~Wallon,
  ``On exotic hybrid meson production in gamma* gamma collisions,''
  Eur.\ Phys.\ J.\ C {\bf 47} (2006) 71
  doi:10.1140/epjc/s2006-02533-7
  [hep-ph/0601176].
  %%CITATION = doi:10.1140/epjc/s2006-02533-7;%%
  %12 citations counted in INSPIRE as of 19 Aug 2019
%\cite{Anikin:2004vc}


%\cite{Teryaev:2001qm}
\bibitem{Teryaev:2001qm}
O.~V.~Teryaev,
``Crossing and radon tomography for generalized parton distributions,''
Phys. Lett. B \textbf{510}, 125-132 (2001)
doi:10.1016/S0370-2693(01)00564-0
[arXiv:hep-ph/0102303 [hep-ph]].
%108 citations counted in INSPIRE as of 14 Feb 2022

  %85 citations counted in INSPIRE as of 31 Oct 2016

\bibitem{Anikin:2004vc}
  I.~V.~Anikin, B.~Pire, L.~Szymanowski, O.~V.~Teryaev and S.~Wallon,
  ``Deep electroproduction of exotic hybrid mesons,''
  Phys.\ Rev.\ D {\bf 70}, 011501 (2004)
  doi:10.1103/PhysRevD.70.011501
  [hep-ph/0401130].
  %%CITATION = doi:10.1103/PhysRevD.70.011501;%%
  %32 citations counted in INSPIRE as of 19 Aug 2019

%\cite{Trachenko:2020ktm}
\bibitem{Trachenko:2020ktm}
K.~Trachenko, V.~Brazhkin and M.~Baggioli,
%``Similarity between the kinematic viscosity of quark-gluon plasma and liquids at the viscosity minimum,''
SciPost Phys. \textbf{10}, no.5, 118 (2021)
doi:10.21468/SciPostPhys.10.5.118
[arXiv:2003.13506 [hep-th]].
%9 citations counted in INSPIRE as of 14 Feb 2022






%\cite{Arbuzov:2020cqg}
\bibitem{Arbuzov:2020cqg}
A.~Arbuzov, A.~Bacchetta, M.~Butenschoen, F.~G.~Celiberto, U.~D'Alesio, M.~Deka, I.~Denisenko, M.~G.~Echevarria, A.~Efremov and N.~Y.~Ivanov, \textit{et al.}
%``On the physics potential to study the gluon content of proton and deuteron at NICA SPD,''
Prog. Part. Nucl. Phys. \textbf{119}, 103858 (2021)
doi:10.1016/j.ppnp.2021.103858
[arXiv:2011.15005 [hep-ex]].
%31 citations counted in INSPIRE as of 14 Feb 2022






















\end{thebibliography}
\end{document}